\documentclass[pra,onecolumn,preprint, 11pt]{revtex4}

\usepackage{amsmath}
\usepackage{graphicx}
\usepackage{dcolumn}
\usepackage{amsfonts}

\begin{document}

\titlepage
\title{Two-Pulse Propagation in Media with Quantum-Mixed Ground States}
\author{B.D. Clader and J.H. Eberly}
\affiliation{Department of Physics and Astronomy \\ University of Rochester \\ Rochester, NY 14627}

\begin{abstract}
We examine fully coherent two-pulse propagation in a lambda-type medium, under two-photon resonance conditions and including inhomogeneous broadening. We examine both the effects of short pulse preparation and the effects of medium preparation. We contrast cases in which the two pulses have matched envelopes or not, and contrast cases in which ground state coherence is present or not.   We find that an extended interpretation of the Area Theorem for single-pulse self-induced transparency (SIT) is able to unify two-pulse propagation scenarios, including some aspects of electromagnetically-induced transparency (EIT) and stimulated Raman scattering (SRS). We present  numerical solutions of both three-level and adiabatically reduced two-level density matrix equations and Maxwell's equations, and show that many features of the solutions are quickly interpreted with the aid of analytic solutions that we also provide for restricted cases of pulse shapes and preparation of the medium. In the limit of large one-photon detuning, we show that the two-level equations commonly used are not reliable for pulse Areas in the 2$\pi$ range, which allows puzzling features of previous numerical work to be understood.
\end{abstract}

\maketitle

\section{Introduction}

The study of the fully coherent propagation of light in the quantum domain, where the non-classical response of a resonant medium is taken into account, was opened by the famous work of McCall and Hahn \cite{mccall-hahn}. They used low-temperature ruby and nanosecond-scale pulses, and reported experimental realization of the first optical solitons, in agreement with their novel theory of self-induced transparency (SIT). McCall and Hahn supplied the theoretical foundation for the SIT effect by self-consistently combining a two-level model for the interacting Cr$^{3+}$ impurity ions with a semiclassical treatment of the field  \cite{Davis}.  The understanding of coherent optical pulse propagation in two-level resonant media was rapidly extended \cite{gibbs-slusher1, gibbs-slusher2, Icsevgi-Lamb, Lamb, allen-eberly}. 

Our interest here is in coherent two-pulse propagation (for an early example, see \cite{Konopnicki-Eberly}). Many different parameters of both the medium and the pulses themselves can play physically distinct roles during propagation. We adopt common restrictions made in treating lambda-type media in order to provide a close focus on a few interesting questions. For example, we assume that the Rotating Wave Approximation is reliable and that the detunings of the two pulses are large and are also equal, so that two-photon resonance applies. 

To focus attention on coherent evolution we also assume that the pulse durations are short enough to ignore homogeneous relaxation of the atoms, but we allow inhomogeneous broadening. To restrict the range of atomic effects we assume that the pulse is too weak to cause excitation beyond the three lambda levels, that the ``total" pulse Area is of order $2\pi$, and that the propagation path is long enough to allow asymptotic conditions of propagation to be reached. A standard sketch is shown in Fig. \ref{lambda}. Because of our focus on fully coherent propagation there are natural connections to SIT, and because our medium is of lambda type, some connections are expected with Harris's electromagnetically-induced transparency (EIT) \cite{Harris} and  traditional stimulated Raman scattering (SRS) \cite{Raman, yariv, milonni-eberly88, bloembergen-wang} and fully coherent ``transient" SRS \cite{transientSRS}, all of which usually employ a quasi-{\em cw} or undepleted pump field instead of a short pump pulse. 

We first present a class of analytic solutions to our two-pulse nonlinear evolution equations (three-level Maxwell-Bloch equations). The solutions support our earlier observation \cite{clader-eberly07} that in the situations of interest one can divide the evolution process into three widely separated propagation zones: a strong interaction or ``pulse transfer" zone, and the zones in the distant past and distant future of the transfer zone. The analytic solutions are analogs of previously known resonant SRS solitons presented by Bol'shov, et al. \cite{bolshov-etal} and our use of the Park-Shin B\"acklund solution approach \cite{park-shin} allows incorporation of far off-resonant pulses and a medium with either pure or quantum-mixed ground states. We find that our solutions are constrained by a new two-pulse Area condition similar to the $2\pi$ Area condition in SIT.

The analytic solutions are then used in several ways. We have found in numerical modeling that the solution formula and the new two-pulse Area condition serve as accurate predictors of output pulse shape and total Area, even for a variety of input conditions not conforming to the analytic solutions. This is closely reminiscent of the behavior of coherent short pulses in one-photon resonant two-level media, which evolve coherently toward the $2\pi$ $sech$ pulse of SIT, even if injected into a medium with shape and Area not in that form.  

We also make use of the analytic solution formulas to check departures from adiabatic predictions. An adiabatic restriction is normal in time-dependent treatments of EIT effects, and we are interested in its limits of applicability in the context of propagation. Similar questions can be asked about SRS, commonly treated as an adiabatic propagation process governed by approximate far off-resonant conditions, where a large product of detuning and pulse duration, $\tau\Delta \gg 1$, is used to justify an approximate two-level set of evolution equations. We  solve numerically for the evolution behavior in the large-detuning case, and demonstrate the inadequacy of the reduced-equation approach if either pulse has near-unit Area. 

In some early numerical work with two-pulse evolution in three-level media \cite{tanno}, a striking and unexplained pulse breakup was reported for solutions of the reduced two-level equations. Our results, obtained in a similar physical situation but with the more fundamental three-level equations, suggest a resolution of this effect, as pulse breakup related to higher order SIT-type multi-pulse formation.  

Finally, the analytic solutions permit a new kind of comment on two-pulse propagation in lambda media. We use them, along with numerical solutions, to show a strong contrast with the earlier reports by Kozlov and Eberly \cite{Kozlov-Eberly} of relative pulse stability of EIT-type and SIT-type propagation behavior. We show here that the presence or absence of quantum coherence of the ground-level preparation of the medium is the crucial property in determining asymptotic stability, and we conclude that mixed-state preparation leads to SIT-type stability, opposite to the conclusion reached for pure-state preparation.


\section{Theoretical Model}

\begin{figure}[h]
\begin{center}
\leavevmode
\includegraphics{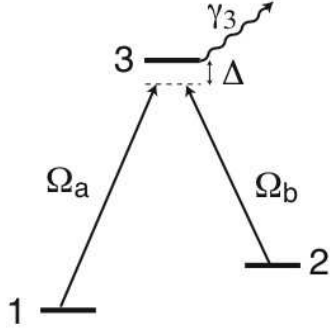}
\end{center}
\caption{\label{lambda} \footnotesize{Three-level atom in two-photon 
resonance with level $1$ connected to level $3$ via the Rabi 
frequency $\Omega_{a}$, referred to as the pump field, and level $2$ 
connected to level $3$ via the Rabi frequency $\Omega_{b}$, referred 
to as the Stokes field.  Each laser field is detuned from the excited 
state resonance by an equal amount $\Delta$, and the excited state 
decay rate is given by $\gamma_3$.}} 
\end{figure}

We consider dual-pulse propagation in a medium of three-level atoms in the lambda configuration as shown in Fig. 
\ref{lambda}.  Given linear polarization, the electric fields of the individual laser pulses can 
be written as scalars: $E_a(x,t) = \mathcal{E}_a(x,t)e^{-i(k_a x - \omega_a 
t)} + \rm{ c.c.}$, where $\mathcal{E}_a$ is the slowly varying 
envelope of the electric field, $k_a$ is the wavenumber, and 
$\omega_a$ is the frequency, with a similar formulation for the 
second pulse $E_b(x,t)$. For the 1-3 transition we have the Rabi frequency $\Omega_a = 
2d_a\mathcal{E}_a/\hbar$, where $d_a$ is the interacting component of the transition dipole moment, and the detuning is $\Delta = \omega_3 - (\omega_1 + \omega_a)$. We use 
corresponding notation for the 2-3 transition. Note that the same 
$\Delta$ serves as the one-photon detuning of each transition below 
resonance, which implies exact two-photon resonance.

The Hamiltonian of the system in the rotating wave picture is given by
\begin{eqnarray} \label{Hamiltonian}
H &=& \hbar\Delta |3\rangle\langle 3| -\hbar\frac{\Omega_{a}}{2}|1\rangle\langle 3| - 
\hbar\frac{\Omega_{b}}{2}|2\rangle\langle 3| 
- \hbar\frac{\Omega_{a}^*}{2}|3\rangle\langle 1| - 
\hbar\frac{\Omega_{b}^*}{2}|3\rangle\langle 2| \nonumber \\
&=& \hbar\Delta |3\rangle\langle 3| -\Big(\hbar\frac{\Omega_{a}}{2}|1\rangle + \hbar\frac{\Omega_{b}}{2}|2\rangle \Big)\langle 3| 
- |3\rangle \Big(\hbar\frac{\Omega_{a}^*}{2}\langle 1| + 
\hbar\frac{\Omega_{b}^*}{2}\langle 2|\Big), 
\end{eqnarray}
where the factorization in the last line suggests introducing the ``bright" and ``dark" state representations  of the two lower states $|1\rangle$ and $|2\rangle$:
\begin{equation} \label{BrightDarkDefs}
|B\rangle \equiv \hbar\frac{\Omega_{a}}{2}|1\rangle + \hbar\frac{\Omega_{b}}{2}|2\rangle \quad {\rm and}\quad |D\rangle \equiv \hbar\frac{\Omega_{b}}{2}|1\rangle - \hbar\frac{\Omega_{a}}{2}|2\rangle,
\end{equation}
where the interaction terms in $H$ clearly depend only on $|B\rangle$, while the orthognal ``dark" state $|D\rangle$ \cite{Arimondo} does not participate in the temporal dynamics. 

Dynamical evolution is confined to $|B\rangle$ and $|3\rangle$ only so long as spatial evolution is ignored. When spatial evolution is permitted, and the dark state is a participant, it can be difficult to maintain some of the restrictions that are commonly imposed in treating two-field physics (fixed control field intensity, a permanently weak probe field, etc.). Some effects of dark-state participation in propagation have been pointed out previously (e.g., see \cite{Eberly-etal94}), and Kozlov and Eberly more recently showed that the dark state even completely controls evolution asymptotically \cite{Kozlov-Eberly, Eberly-Kozlov} when propagation occurs in a lambda medium of two coherently prepared ground states (sometimes called ``phaseonium" \cite{Scully}). The contrast extends to stimulated Raman scattering, historically the first lambda-medium propagation effect, and  not associated with an extended prepared coherence between the two ground levels.

We now derive analytically exact two-pulse solutions for lambda media. This requires an extension of previous soliton theories for resonant SRS propagation (see \cite{bolshov-etal, park-shin}). Here we depart substantially from resonance, however still near enough that we can neglect anti-Stokes effects or interaction with any other excited states.  We adopt SRS nomenclature and will refer to $\Omega_a$ as the pump field and $\Omega_b$ as the Stokes field.  We will focus on non-phaseonium medium preparation, but include a variety of  mixed-state descriptions of levels 1 and 2. This permits us to account for incoherent initial population distributions that may arise from thermal effects, incoherent optical pumping of the ground states, etc.  The solution method we use allows for significant changes in both the pump and Stokes pulses, and gives analytical insight into previously pursued numerical studies of short-pulse Raman propagation \cite{tanno}.

The von Neumann equation for the atomic density matrix is
\begin{equation}
\label{DensityMatrix}
i\hbar\dot{\rho} = [H,\rho],
\end{equation}
where the over-dot signifies $\partial/\partial t$ and where $H$ is given in Eq. (\ref{Hamiltonian}). For later reference, the equations for the density matrix elements are:
\begin{subequations} 
\label{rhoEquations}
\begin{align}
\dot\rho_{11} =& i\frac{\Omega_a}{2}\rho_{31} - i\frac{\Omega_a^*}{2}\rho_{13} \\
\dot\rho_{22} =& i\frac{\Omega_b}{2}\rho_{32} - i\frac{\Omega_b^*}{2}\rho_{23} \\
\dot\rho_{33} =& -i\frac{\Omega_a}{2}\rho_{31} + i\frac{\Omega_a^*}{2}\rho_{13}
- i\frac{\Omega_b}{2}\rho_{32} + i\frac{\Omega_b^*}{2}\rho_{23} \\
\dot\rho_{12} =&  i\frac{\Omega_a}{2}\rho_{32} -i\frac{\Omega_b^*}{2}\rho_{13} \\
\dot\rho_{13} =& i\Delta \rho_{13} - i\frac{\Omega_b}{2}\rho_{12} + i\frac{\Omega_a}{2}(\rho_{33} - \rho_{11}) \\
\dot\rho_{23} =& i\Delta \rho_{23} - i\frac{\Omega_a}{2}\rho_{21} + i\frac{\Omega_b}{2}(\rho_{33} - \rho_{22}). 
\end{align}
\end{subequations}
Without damping terms, Eqs. \eqref{DensityMatrix} and \eqref{rhoEquations} imply that the duration of the laser pulses under consideration is much shorter than all homogeneous relaxation processes (such as the excited state decay, with rate $\gamma_3$, shown in Fig. \ref{lambda}), thus justifying their neglect.

For simplicity we will change to a travelling reference frame where the time derivatives are with respect to the retarded time $T=t-x/c$ and the spatial derivatives are with respect to $Z=x/c$.   This specification gives
\begin{equation} \label{ZTdefs}
\frac{\partial}{\partial t} = \frac{\partial}{\partial T} \quad {\rm and} \quad c\frac{\partial}{\partial x} + \frac{\partial}{\partial t} = 
\frac{\partial}{\partial Z}.
\end{equation}

The field evolves according to Maxwell's equations.  In effect, the interaction Hamiltonian itself, which is just the dipole interaction $\vec d \cdot \vec E(x,t)$, evolves according to Maxwell's equations. We assume that the pump field only interacts with the $1\to3$ transition, the Stokes field only interacts with the $2\to3$ transition, and the fields move only in the positive $x$ direction.  These assumptions are justified by the inequality $|(\omega_2-\omega_1)/\Delta| \gg 1$ together with the slowly varying envelope approximation, and they give two distinct field equations from Maxwell's equations:
\begin{subequations}
\label{MaxwellEquation}
\begin{align}
\frac{\partial \Omega_{a}}{\partial Z} & = -i\mu_{a}\rho_{13} \\
\frac{\partial \Omega_{b}}{\partial Z} & = -i\mu_{b}\rho_{23}.
\end{align}
\end{subequations}
Here $N$ is the density of atoms, $\mu_a=Nd_a\omega_a^2/\hbar\epsilon_0$ is the atom field coupling parameter for the $1 \to 3$ transition, and similarly $\mu_b$ for transition $2 \to 3$. For simplicity in the derivations to be presented, we have not included inhomogeneous broadening in these equations, but it will be incorporated in the end.  We note that when $\mu_a=\mu_b \equiv \mu$ these results can be combined into a single matrix equation for the spatial evolution of the Hamiltonian:
\begin{equation}
\label{HamMaxEqn}
\frac{\partial H}{\partial Z} = -\frac{\hbar\mu}{2}[W,\rho],
\end{equation}
where
\begin{equation}
\label{WMatrix}
W = -\frac{i}{2}
\begin{pmatrix}
1 & 0 & 0\\
0 & 1 & 0\\
0 & 0 & -1
\end{pmatrix},
\end{equation}
and this matrix format will be useful shortly.  For notational simplicity we have suppressed the functional dependences of the density matrix elements in Eq. \eqref{DensityMatrix} and the Rabi frequencies in Eqns. \eqref{MaxwellEquation} on $T$ and $Z$ (or equivalently $x$ and $t$), and we will do so for the remainder of the paper.


\section{Park-Shin B\"{a}cklund Solution Method}\label{ss:backlund}

Eqns. \eqref{DensityMatrix} and \eqref{MaxwellEquation} are known as the Maxwell-Bloch equations for the three-level lambda system.  For comparison with adiabatically approximated solution methods as well as with numerical methods it is highly useful to have solutions both in analytic form and without adiabatic assumptions.  The inverse scattering technique has previously been used to construct resonant solutions \cite{bolshov-etal}.  These equations are integrable when $\mu_a = \mu_b \equiv \mu$, which we will assume from now on.  We will solve these equations using the Park-Shin (PS) formulation \cite{park-shin} of the B\"acklund solution method.  We will briefly sketch the PS method, and slightly generalize their approach to include arbitrary detuning of fields.

One begins by finding a pair of linear operators $L_1$ and $L_2$, which will allow for a linear representation of the nonlinear Maxwell-Bloch equations when subject to the constraint that 
$[L_1,L_2]=0$.  We define these operators as:
\begin{subequations}
\label{L}
\begin{align}
L_1 & =  \frac{\partial}{\partial T}+ U +\lambda W  \label{L1} \\
L_2 & = (\lambda-\Delta )\frac{\partial}{\partial Z} + V \label{L2},
\end{align}
\end{subequations}
where $\lambda$ is the spectral parameter, and $U$ and $V$ are given by
\begin{equation}
\label{Ulax}
U = \frac{i}{\hbar}H^0
\end{equation}
\begin{equation}
\label{Vlax}
V = -\frac{i\mu}{2}\rho,
\end{equation}
where $H^0$ is the Hamiltonian in Eq. \eqref{Hamiltonian} with 
$\Delta=0$.  To recover the Maxwell-Bloch equations one imposes the commutator 
relation $[L_1,L_2] = 0$ and then compares terms with like orders of 
$\lambda$.   In doing so one finds the two independent $3 \times 3$ 
matrix equations
\begin{subequations}
\begin{align}
i\hbar\frac{\partial \rho}{\partial T} = [H,\rho] \label{schr-UV} \\
\frac{\partial U}{\partial Z} = [W,V] \label{maxwell-UV}.
\end{align}
\end{subequations}
Eq. \eqref{schr-UV} is simply the von Neumann equation, and it is straightforward to show that Eq. \eqref{maxwell-UV} is equivalent to the matrix form of Maxwell's equation presented in Eq. \eqref{HamMaxEqn}.  Thus we see that the commutator $[L_1,L_2]=0$ correctly reproduces the Maxwell-Bloch equations, and is equivalent to the integrability condition, $\frac{\partial V^\prime}{\partial T} - \frac{\partial U^\prime}{\partial Z} + [V^\prime, U^\prime] = 0$, of the ``Lax" Pair $U^\prime = -(U+\lambda W)$ and $V^\prime = -\frac{1}{\lambda - \Delta} V$.

The PS method now proceeds as follows.  Because $L_1$ and $L_2$ commute, they have common eigenvectors, which we represent here in the form of matrix $M$: 
\begin{subequations}
\label{L_M}
\begin{align}
L_1M & =  0 \\
L_2M & = 0.
\end{align}
\end{subequations}
We now apply the B\"acklund transformation to Eqns. \eqref{L_M} by 
writing the solution matrix $M$ in terms of a known ``proto-solution" matrix, which we denote as $M^{(0)}$, and a dressing operator $\Phi: ~M 
\equiv \Phi M^{(0)}$. The proto-solution can even be a trivial 
solution. We will write $\Phi$ as
\begin{equation}
\label{zsdressing}
\Phi = 1+i\frac{\eta}{\lambda}(2P-1),
\end{equation}
where $P$ is a projection operator in the $3\times 3$ atomic space, implying the existence of a vector $|s\rangle$ such that 
\begin{equation}
\label{proj-op}
P = \frac{|s\rangle\langle s |}{\langle s|s\rangle},
\end{equation}
and $\eta$ is a parameter of the B\"acklund transformation, which we take to be 
real. Eqns. \eqref{L_M} and \eqref{zsdressing} combine to give:
\begin{subequations}
\label{Phi_Eq}
\begin{align}
\frac{\partial}{\partial T}\Phi -\Phi(U^{(0)}+\lambda W)+(U+\lambda 
W)\Phi=0 \label{transf1}
\\
(\lambda-\Delta)\frac{\partial}{\partial Z}\Phi+V\Phi-\Phi V^{(0)}=0 
\label{transf2},
\end{align}
\end{subequations}
where $U^{(0)}$ and $V^{(0)}$ are the known solutions of the fields and density matrix making up the proto-solution matrix $M^{(0)}$.  We now insert the definition of $\Phi$ into Eqns. \eqref{Phi_Eq}, and compare terms with like orders of $\lambda$, which results in four independent equations. This allows one to eliminate equations involving the space and time derivatives to arrive at the following two equations:
\begin{subequations}
\label{UV-Sol}
\begin{align}
U & = U^{(0)}-2i\eta [W,P] \label{solfield}
\\
V & = \frac{\eta^2}{\eta^2 + 
\Delta^2}\left(2P-1-i\frac{\Delta}{\eta}\right)V^{(0)}\left(2P-1+i\frac{\Delta}{\eta}\right) 
\label{solrho}.
\end{align}
\end{subequations}
Eqns. \eqref{UV-Sol} relate the unknown solutions given by $U$ and $V$ to the known solutions given by $U^{(0)}$ and $V^{(0)}$ and the matrix $P$.

The only unknown quantity in Eqns. \eqref{UV-Sol} is the projection operator $P$ (or the vector $|s\rangle$).  We can find an equation for $P$ in terms of known quantities by again using Eqns. \eqref{Phi_Eq}, but this time solving for terms containing the derivatives.  The result, after simplifications made possible by utilizing projection operator properties, is two independent linear differential equations for the vector $|s\rangle$, given by:
\begin{subequations}
\begin{align}
\left(\frac{\partial}{\partial T} + U^{(0)}  - 
i\eta W\right)|s\rangle=0 \label{fin1}
\\
\left(\frac{\partial}{\partial 
Z}-\frac{\Delta-i\eta}{\eta^2+\Delta^2}V^{(0)}\right)|s\rangle=0 
\label{fin2}.
\end{align}
\end{subequations}
Since the commutator $[U ^{(0)} - i\eta W,V ^{(0)} ] $ vanishes, the solutions to Eqns. \eqref{fin1} and \eqref{fin2} can be exponentiated together to give
\begin{equation}
\label{sols}
|s\rangle=\textnormal{exp}\left(-U ^{(0)}T  +  i\eta WT + 
\frac{\Delta -i\eta}{\eta^2+\Delta^2}V ^{(0)} 
Z\right)|u\rangle,
\end{equation}
where $|u\rangle$ is an arbitrary constant vector, with coefficients 
$(u_1,u_2,u_3)$, which is specified by the initial condition of the 
pulse.  At this point, one can introduce inhomogeneous broadening by 
simply Doppler averaging the third term in Eq. \eqref{sols} with the 
Doppler distribution function
\begin{equation} \label{inhomogDef}
F(\Delta) \equiv \frac{T_2^*}{\sqrt{2\pi}} 
e^{-(\Delta-\bar{\Delta})^2(T_2^*)^2/2},
\end{equation}
where $T_2^*$ is the inhomogeneous lifetime and $\bar{\Delta}$ is the 
line-center single-photon detuning of both laser fields.  To be consistent one must also average the right hand side of Maxwell's equation \eqref{MaxwellEquation} with the same function.  This could have been incorporated from the start.

Finally, since we are interested in the actual density matrix $\rho$ 
and the fields $\Omega_a$ and $\Omega_b$ (instead of $U$ and $V$), we 
write them in terms of the vector $|s\rangle$ using Eqns. 
\eqref{solrho} and \eqref{solfield}.  The density matrix solution is:
\begin{equation}
\label{solDens}
\rho =  \frac{\eta^2}{\eta^2 + 
\Delta^2}\left(2\frac{|s\rangle\langle s|}{\langle 
s|s\rangle}-\frac{\eta+i\Delta}{\eta}\right)\rho^{(0)}\left(2\frac{|s\rangle\langle 
s|}{\langle s|s\rangle}-\frac{\eta-i\Delta}{\eta}\right),
\end{equation}
and the individual field solutions are:
\begin{subequations}
\label{solOmega}
\begin{align}
\Omega_a & = 2i\langle 1|U|3\rangle = \Omega_a^{(0)} - 
4i\eta\frac{\langle 1|s\rangle\langle s|3\rangle}{\langle s|s\rangle} 
\\
\Omega_b & = 2i\langle 2|U|3\rangle = \Omega_b^{(0)} - 
4i\eta\frac{\langle 2|s\rangle\langle s|3\rangle}{\langle 
s|s\rangle}
\end{align}
\end{subequations}
To summarize, the coupled non-linear Maxwell-Bloch equations \eqref{DensityMatrix} and \eqref{MaxwellEquation} have been re-written as two coupled linear differential equations \eqref{fin1} and \eqref{fin2} and their solution is given in equation \eqref{sols}.  The solution depends on a known 
proto-solution to the original coupled Maxwell-Bloch equations given by $U^{(0)}$ and $V^{(0)}$, or equivalently $\Omega_a^{(0)}$, $\Omega_b^{(0)}$, and $\rho^{(0)}$.  Any solution of the Maxwell-Bloch equations can be used as this proto-solution.  In the next section we will show how even trivial proto-solutions will yield interesting results.


\section{Two-Pulse Analytic Solutions} \label{analytical_solutions}
We now use the PS approach to solve the MB equations explicitly for the case off-resonant SRS pulse transfer.  The solution method allows one to treat mixed as well as pure-state media and also allows for treatment of spatially-dependent initial level populations \cite{clader-eberly, clader-eberly2}.  The choice of proto-solution gives the initial condition of the medium and pulses at $T=-\infty$.

The state of a lambda atom with no population in the upper level has the form $|\psi_{\text{atom}}(0)\rangle = c_1|1\rangle + c_2|2\rangle$, so if we write $c_1=\alpha e^{i\phi_1}$, and $c_2=\beta e^{i\phi_2}$, with real $\alpha$ and $\beta$ constrained by $\alpha^2 + \beta^2 = 1$, we can identify phase effects easily in the corresponding density matrix
\begin{equation}
\label{InConditionPure}
\rho(0) = |\psi_{\text{atom}}(0)\rangle\langle\psi_{\text{atom}}(0)| = \begin{pmatrix}
\alpha^2 & \alpha\beta e^{i \delta\phi} & 0 \\
\alpha\beta e^{-i \delta\phi} & \beta^2 & 0 \\
0 & 0 & 0
\end{pmatrix},
\end{equation}
where $\delta\phi = \phi_1-\phi_2$.  Complete initial dephasing is a natural situation consistent, for example, with thermal mixing between the two ground state populations or with incoherent optical pumping between them.  In this case the off-diagonal terms of the density matrix vanish and we have a diagonal initial density matrix:
\begin{equation}
\label{InCondition}
\rho(0) \to \begin{pmatrix}
\alpha^2 & 0 & 0 \\
0 & \beta^2 & 0 \\
0 & 0 & 0
\end{pmatrix}.
\end{equation}

We easily see that the state specification given in Eq. \eqref{InCondition}, along with the zero values $\Omega_a^{(0)} = 0$ and $\Omega_b^{(0)}=0$, constitutes a trivial but exact solution of the MB equations. It is also useful because the PS B\"acklund approach constructs a non-trivial solution from it.  The trivial solutions give $U^{(0)}=0$ and $V^{(0)}=-ig\rho(0)$.  Since these solutions give a diagonal matrix in the exponent of Eq. \eqref{sols}, the components of vector $|s\rangle$ are easily calculated.  They are:
\begin{subequations}
\label{vecScomponents}
\begin{align}
\langle 1|s\rangle & = \exp\bigg(\frac{1}{2\tau}\bigg)\bigg(T - \alpha^2\mu\tau^2\bigg\langle \frac{1}{1+i\Delta\tau}\bigg\rangle Z\bigg)\\
\langle 2|s\rangle & = \exp\bigg(\frac{1}{2\tau}\bigg)\bigg(T - \beta^2\mu\tau^2\bigg\langle \frac{1}{1+i\Delta\tau}\bigg\rangle Z\bigg)\\
\langle 3|s\rangle & = -i \exp\bigg(-\frac{T}{2\tau}\bigg),
\end{align}
\end{subequations}
where we have identified $1/\eta \equiv \tau$ as the nominal pulse width and take $u_1=1$, $u_2=1$, and $u_3=-i$ for greatest simplicity.  Eqns. \eqref{solOmega} and \eqref{vecScomponents} combine to give the two pulse solutions:
\begin{subequations}
\label{PulseSolution}
\begin{align}
\Omega_{a} & = \frac{4e^{-i\alpha^2 \delta Z}}{\tau} \bigg[2 \textnormal{cosh} \big(T/\tau - \alpha^2 \kappa Z\big) + \exp\big(T/\tau + (\alpha^2-2\beta^2) \kappa Z\big) \bigg]^{-1} \\
\Omega_{b} & = \frac{4e^{-i\beta^2 \delta Z}}{\tau}\bigg[2 \textnormal{cosh} \big(T/\tau -\beta^2 \kappa Z\big) + \exp\big(T/\tau +(\beta^2-2\alpha^2) \kappa  Z \big)\bigg]^{-1},
\end{align}
\end{subequations}
where the functions $\kappa$ and $\delta$ are given by the inhomogeneous average:
\begin{equation}
\delta + i\kappa = \frac{\mu}{2}\int\frac{F(\Delta)~d\Delta}{\Delta - i/\tau},
\end{equation}
or
\begin{equation}
\label{lengthScale}
\kappa = \frac{\mu}{2 \tau} \int_{-\infty}^{\infty}\frac{F(\Delta)d\Delta}{\Delta^2 + \left(\frac{1}{\tau} \right)^2} \quad{\rm and}\quad
\delta = \frac{\mu}{2} \int_{-\infty}^{\infty}\frac{\Delta F(\Delta)d\Delta}{\Delta^2 + \left(\frac{1}{\tau} \right)^2}.
\end{equation}
Here $\delta/c$ adds to the index of refraction and $\kappa/c$ is the absorption depth for weak-field excitation and so it sets the distance scale for propagation effects. 

Substantial transient atomic coherence develops during the course of propagation and off-diagonal density matrix elements become non-zero.  These  solutions are calculated from Eqns. \eqref{solDens} and \eqref{vecScomponents}, giving:
\begin{subequations}
\label{DensityMatrixSolution}
\begin{align}
\rho_{11} & = \frac{1}{1+(\Delta\tau)^2}\bigg(\alpha^2(|f_{11}|^2+(\Delta \tau)^2) + \beta^2|f_{12}|^2\bigg) \\
\rho_{22} & = \frac{1}{1+(\Delta\tau)^2}\bigg((\alpha^2|f_{12}|^2 + \beta^2(|f_{22}|^2+(\Delta\tau)^2)\bigg) \\
\rho_{33} & = \frac{1}{1+(\Delta\tau)^2}\bigg((\alpha^2|f_{13}|^2 + \beta^2|f_{23}|^2)\bigg) \label{excited_state}  \\
\rho_{12} & = \frac{1}{1+(\Delta\tau)^2}\bigg((\alpha^2(f_{11}-i\Delta\tau)f_{12} + \beta^2(f_{22}+i\Delta\tau)f_{12})\bigg) \\
\label{Coherence13} \rho_{13} & = \frac{1}{1+(\Delta\tau)^2}\bigg((\alpha^2(f_{11}-i\Delta\tau)f_{13} + \beta^2f_{12}f_{23})\bigg) \\
\label{Coherence23} \rho_{23} & =\frac{1}{1+(\Delta\tau)^2}\bigg( (\alpha^2f_{12}^*f_{13} + \beta^2(f_{22}-i\Delta\tau)f_{23})\bigg),
\end{align}
\end{subequations}
where the space-time dependences are contained in the functions $f_{ij}$:
\begin{subequations}
\label{fFunctions}
\begin{align}
f_{11} & = \bigg[2 \textnormal{ sinh} \big(T/\tau-\alpha^2 \kappa Z\big) - 
\exp\big(T/\tau+(\alpha^2-2\beta^2)\kappa Z \big)\bigg]\bigg/D(Z,T) \\
f_{22} & = \bigg[-2 \textnormal{ cosh }\big(T/\tau-\alpha^2 \kappa Z\big) + 
\exp\big(T/\tau+(\alpha^2-2\beta^2)\kappa Z\big)\bigg]\bigg/D(Z,T) \\
f_{12} & = 2 e^{-i(\alpha^2-\beta^2) \delta Z}e^{T/\tau-\beta^2 \kappa Z}/D(Z,T) \\
f_{13} & = 2 ie^{-i\alpha^2 \delta Z}/D(Z,T) \\
f_{23} & = 2 i e^{-i\beta^2 \delta Z}e^{(\alpha^2-\beta^2) \kappa Z}/D(Z,T),
\end{align}
\end{subequations}
and the denominator function $D(Z,T)$ is given by
\begin{subequations}
\begin{align}
\label{DFunction}
D(Z,T) & = 2 \textnormal{ cosh }\big(T/\tau-\alpha^2 \kappa Z\big) + \exp\big(T/\tau+(\alpha^2-2\beta^2) \kappa Z\big).
\end{align}
\end{subequations}

\subsection{Input Regime: $-\kappa Z \gg 1$}
The general pulse and density matrix solutions given in Eqns. \eqref{PulseSolution} and \eqref{DensityMatrixSolution} are cumbersome and resist rapid interpretation.  However their key asymptotic features are clear.  First, in the far negative spatial region, i.e. $-\kappa Z \gg 1$.  The asymptotic forms of Eqns. \eqref{PulseSolution} simplify to
\begin{subequations}
\label{InputPulseSolution}
\begin{align}
\Omega_{a} & = \frac{2e^{-i\alpha^2 \delta Z}}{\tau}\textnormal{ sech}(T/\tau - \alpha^2 \kappa Z) \\
\Omega_{b} & = 0,
\end{align}
\end{subequations}
which one can easily recognize as the McCall-Hahn $2\pi$ hyperbolic secant pulse solutions \cite{mccall-hahn} for SIT on the $1 \to 3$ transition, and no pulse on the $2 \to 3$ transition.  We interpret this as an ``input"  pulse advancing with a group velocity given by $v_g^{(a)}/c = (1+\alpha^2 \kappa \tau)^{-1}$.  The density matrix solutions also simplify substantially in the same limit.  They are
\begin{subequations}
\label{InputDensityMatrixSolution}
\begin{align}
\rho_{11} & = \frac{1}{1+(\Delta\tau)^2}\alpha^2\big(\tanh^2(T/\tau-\alpha^2 \kappa Z) +(\Delta\tau)^2\big) \\
\rho_{22} & = \beta^2\\
\rho_{33} & = \frac{1}{1+(\Delta\tau)^2}\alpha^2\textnormal{ sech}^2(T/\tau-\alpha^2 \kappa Z)  \\
\rho_{12} & = 0 \\
\rho_{13} & = i\frac{1}{1+(\Delta\tau)^2}\alpha^2e^{-i\alpha^2 \delta Z}\textnormal{ sech}(T/\tau-\alpha^2 \kappa Z)\big(\tanh(T/\tau-\alpha^2 \kappa Z) - i \Delta\tau\big) \\
\rho_{23} & =0.
\end{align}
\end{subequations}
One can see that all of the interaction in this asymptotic limit is localized to the $1 \to 3$ transition, and the three-level solutions simplify to those for a two-level atom.

\subsection{Output Regime: $\kappa Z \gg 1$}
Similarly we also look at the solutions for large positive propagation distance where $\kappa Z \gg 1$, the ``output" regime.  These solutions are given by
\begin{subequations}
\label{OutputPulseSolution}
\begin{align}
\Omega_{a} & = 0 \\
\Omega_{b} & = \frac{2e^{-i\beta^2 \delta Z}}{\tau}\textnormal{ sech}(T/\tau - \beta^2 \kappa Z),
\end{align}
\end{subequations}
which is another SIT solution except that the excitation has been transferred to the Stokes pulse, and here $\Omega_b$ moves with group velocity $v_g^{(b)}/c = (1+\beta^2 \kappa \tau)^{-1}$.  The density matrix solutions similarly change.  They are now given by
\begin{subequations}
\label{OutputDensityMatrixSolution}
\begin{align}
\rho_{11} & = \alpha^2 \\
\rho_{22} & =\frac{1}{1+(\Delta\tau)^2}\beta^2\big(\tanh^2(T/\tau-\beta^2 \kappa Z) +(\Delta\tau)^2\big) \\
\rho_{33} & = \frac{1}{1+(\Delta\tau)^2}\beta^2\textnormal{ sech}^2(T/\tau-\beta^2 \kappa Z)  \\
\rho_{12} & = 0 \\
\rho_{13} & = 0 \\
\rho_{23} & = i\frac{1}{1+(\Delta\tau)^2}\beta^2e^{-i\beta^2 \delta Z}\textnormal{ sech}(T/\tau-\beta^2 \kappa Z)\big(\tanh(T/\tau-\beta^2 \kappa Z) - i \Delta\tau\big).
\end{align}
\end{subequations}

\subsection{Pure State Input/Output Regimes}
As a limiting case we take the initial preparation of the atoms to be in a pure-state by setting $\alpha = 1$ and $\beta=0$.  In this case the input pulses for $-\kappa Z \gg 1$ are given 
\begin{subequations}
\label{InputPulseSolutionPureState}
\begin{align}
\Omega_{a} & = \frac{2e^{-i \delta Z}}{\tau}\textnormal{ sech}(T/\tau - \kappa Z) \\
\Omega_{b} & = 0,
\end{align}
\end{subequations}
while the corresponding input regime excited state probability is
\begin{equation}
\label{excitedStateInput}
\rho_{33} = \frac{1}{1+(\Delta\tau)^2}\text{sech }^2\bigg(\frac{T}{\tau} - \kappa Z\bigg).
\end{equation}
The output pulses for $\kappa Z \gg 1$ are similarly found and given by
 \begin{subequations}
\label{OutputPulseSolutionPureState}
\begin{align}
\Omega_{a} & = 0 \\
\Omega_{b} & = \frac{2}{\tau}\textnormal{ sech}(T/\tau),
\end{align}
\end{subequations}
and the output regime the excited state probability is simply
\begin{equation}
\label{excitedStateOutput}
\rho_{33} = 0.
\end{equation}
One can verify these forms by comparing them to the mixed-state case, taking $\alpha = 1$ and $\beta = 0$, and applying the appropriate limits for the input and output regimes.

In this limiting case the input pump pulse is identical in form to an SIT pulse and is coupled to the $1\to3$ absorbing transition.  It causes coherent atom excitation, and travels with reduced group velocity $v_g^{(a)}/c = (1+\kappa \tau)^{-1}$.  Meanwhile the output Stokes pulse, while similar in temporal shape to an SIT pulse is completely decoupled from the medium, and thus moves with group velocity $v_g^{(b)}/c = 1$.  This is due to the fact that we have neglected any anti-Stokes interaction, and have limited our atomic system to three levels.  This is a trivial example of a Dark State where lack of population in state 2, rather than quantum interference effects, causes the medium to be transparent to the Stokes pulse.

\section{Pulse Area and Transfer Zone Behavior}

\begin{figure}[h]
\includegraphics[height=2in]{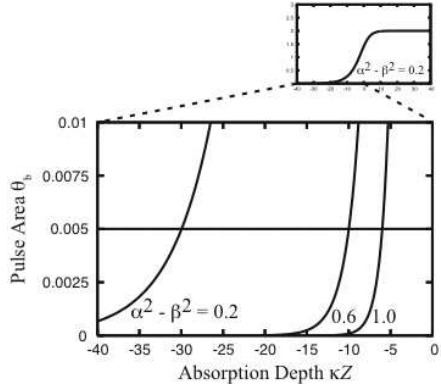}
\caption{\label{fig.AreaEvol} Curves showing the way the analytic formula (\ref{PulseAreas}) predicts $\theta_b$ to grow toward  $Z=0$ where its Area matches the pump pulse Area $\theta_a$. The distance required for this growth from the initial value $\theta_b = 0.005 \pi$ depends on the so-called Raman inversion $\alpha^2 - \beta^2$, and three examples are given as indicated to the right of each graph. These distances are designated as $Z_T$ in the text, to indicate the distance of propagation in the medium required to reach the ``transfer" point where $\theta_b = \theta_a$.   The large plot shows the early growth stages for very small values of $\theta_b$ while the small plot shows the entire growth of the Stokes pulse.}
\end{figure}

Since the solutions in both forward and backward asymptotic regimes reduce to two-level solitons, in these regimes the pulses are expected to obey the two-level Area Theorem \cite{mccall-hahn}. The Area of a pulse is defined to be
\begin{equation}
\label{pulseArea}
\theta(Z) = \int_{-\infty}^{\infty}\Omega(Z,T)dT,
\end{equation}
from which the pulse Areas of solutions \eqref{PulseSolution} can be shown to be
\begin{equation}
\label{PulseAreas}
\theta_{a}(Z) = \frac{2\pi}{h(Z)}\quad {\rm and} \quad \theta_{b}(Z) =\frac{2\pi}{h(-Z)}
\end{equation}
where
\begin{equation}
\label{AreaFunction}
h(Z) = \sqrt{1+e^{2(\alpha^2-\beta^2)\kappa Z}}.
\end{equation}
The depth of medium required for significant change in pulse Area obviously depends on the so-called ``Raman inversion" $\alpha^2 - \beta^2$, and this is indcated in the plots of $\theta_b(Z)$ in Fig. \ref{fig.AreaEvol}.

From the analytic formulas we derived one easily finds the $\mp Z$ asymptotic limits:  $\theta_{a}(Z) \to 2\pi$ and $\theta_{b}(Z) \to 2\pi$, as noted in the previous Section. However, in addition, one can extract a surprising new result if one defines a ``total" Area $\theta(Z)$ for any position $Z$:
\begin{equation}
\label{totalArea}
\theta(Z) \equiv \sqrt{\theta_a^2(Z) + \theta_b^2(Z)} = 2\pi.
\end{equation}
This makes the connection to SIT very close indeed, and shows unexpected continuity of total Area through the non-SIT transfer zone.  The significance of this exact result for us is that the pulse strengths and pulse durations of the exact solutions are fully linked through the two-pulse Eqn. (\ref{totalArea}). This gives the impression that the analytic solutions apply only to situations in which at least one pulse is ``strong", with an Area greater than $\pi$. We will comment further on this point in the next Section.

Another point is that our solutions describe a single-peak pump pulse and a single-peak Stokes pulse.  For larger total pulse Areas (e.g., $3\pi < \theta < 5\pi$) one would expect to obtain double-pulse soliton solutions.  We do not consider higher than single soliton solutions here, but they can be obtained by using a nonlinear superposition rule that can be derived from the B\"acklund transformation.  For more we refer the reader to \cite{park-shin}.

We have already reported \cite{clader-eberly07} that the analytic solutions describe the transfer of a $2\pi$ Area sech-shaped pulse on the $1 \to 3$ transition to a $2\pi$ Area sech-shaped pulse on the $2 \to 3$ transition, through an intermediate interaction regime or ``transfer zone".  We can obviously use this three-regime language here, where regimes I and III contain the asymptotic input and output pulses described by Eqns. \eqref{InputPulseSolution}  - \eqref{OutputDensityMatrixSolution}, and regime II contains the transfer between them, where the full three level equations \eqref{PulseSolution} and \eqref{DensityMatrixSolution} are needed.  Regime II can be thought of as exhibiting a strong-field SRS process.  

\begin{figure}[h]
\begin{center}
\leavevmode
\includegraphics{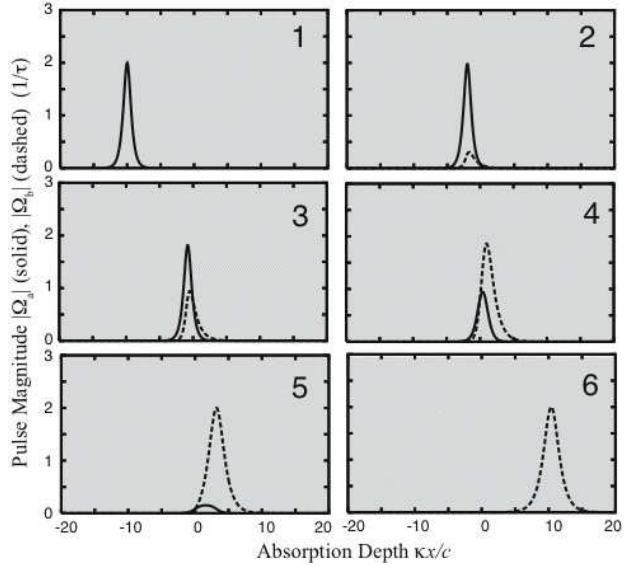}
\end{center}
\caption{\label{fig.pulse} Plots of the analytic pulse solutions given in Eq. \eqref{PulseSolution}.  The horizontal axis is $x$ in units of $\kappa/c$, and the vertical axis is the pulse Rabi frequency in units of $\tau^{-1}$. The background is slightly shaded to indicate the presence of the lambda medium.  The solid curve is the pump pulse, $|\Omega_{a}|$, and the dashed curve is the Stokes pulse, $|\Omega_{b}|$.  The plot shows Raman amplification with complete pump depletion and Stokes amplification.  The input pulse on the 1 - 3 transition amplifies a weak probe pulse on the 2 - 3 transition. Parameters: $\alpha^2 = 1.0$, $\beta^2 = 0.0$, $\tau \approx 3T_2^*$, and $\bar{\Delta}\tau=10$.}
\end{figure}

The analytic solutions for an off-resonant interaction (i.e. $\bar{\Delta}\tau=10$) are shown in Fig. \ref{fig.pulse}, and the three-regime behavior just discussed is clearly evident.  The pulses are plotted as a function of the laboratory frame position variable $x$ in units of $\kappa/c$, and each frame corresponds to a time $t$ in units of $\tau$.  As stated previously these are related to the coordinates presented in the solution by $T=t-x/c$ and $Z=x/c$.  In these examples, during the input stage of propagation, only the pump pulse is significantly present,  corresponding to regime I (frames 1 and 2 of Fig. \ref{fig.pulse}).  During ``transfer" from pump to Stokes both pulses are intense, indicating that we are in regime II (frames 3-5).  Finally the asymptotic output emerges and only the Stokes pulse is significantly present (frame 6) indicating we have reached regime III.


\section{Flux Conservation}
It may appear ``obvious" that energy conservation cannot be satisfied through the Raman exchange regime II.  If the original input pulse is composed entirely of type ``a" photons as seen in Eqns. \eqref{InputPulseSolutionPureState}, while the output pulse is composed of different frequency type ``b" photons as shown in Eqns. \eqref{OutputPulseSolutionPureState},  the energy difference associated with the different pulse frequencies appears to be lost, since the medium stores no energy at the end.

In the situation at hand the interacting-system conservation law is one of flux, not of energy.  An expression for Poynting's Theorem in this one-dimensional example can be derived from the coupled Maxwell-Bloch equations \eqref{rhoEquations} and \eqref{MaxwellEquation} and is given by

\begin{equation}
\label{PoyntingsTheorem}
\frac{\partial}{\partial Z}(|\Omega_a|^2 + |\Omega_b|^2)  + 2\mu\frac{\partial\langle \rho_{33}\rangle}{\partial T} = 0.
\end{equation}
In the input regime I where only a pump pulse is present, Poynting's Theorem becomes
\begin{equation}
\label{PoyntingsTheoremInput}
\frac{\partial}{\partial Z}|\Omega_a|^2 + 2\mu\frac{\partial\langle \rho_{33}\rangle}{\partial T} = 0,
\end{equation}
and for the output regime III where only the Stokes pulse is present we get
\begin{equation}
\label{PoyntingsTheoremOutput}
\frac{\partial}{\partial Z} |\Omega_b|^2 = 0,
\end{equation}
where the $\rho_{33}$ term also vanishes as shown in Eq. \eqref{excitedStateOutput} since we are considering the pure-state for this example.

The regime III Stokes pulse solution, given in Eq.  \eqref{OutputPulseSolutionPureState}, trivially satisfies Eq. \eqref{PoyntingsTheoremOutput}, since it is completely decoupled from the medium and thus has no $Z$ dependence.  To view in the laboratory space-time variables, we recall the definitions of $Z$ and $T$ defined in Eq. \eqref{ZTdefs} and write Eq. \eqref{PoyntingsTheoremOutput} as
\begin{equation}
\label{PoyntingOutputLabST}
\frac{\partial}{\partial Z} |\Omega_b|^2 = \frac{4}{\tau^2}\bigg(c\frac{\partial}{\partial x} + \frac{\partial}{\partial t}\bigg)\text{sech}^2\bigg(\frac{t-x/c}{\tau}\bigg) = 0.
\end{equation}
Viewed in these variables we see that Poyting's Theorem is still obviously satisfied and that the Stokes pulse is moving at the speed of light, due to the fact that it is decoupled from the medium.

To verify Poynting's Theorem in regime I, we insert the pulse solutions from Eqns. \eqref{InputPulseSolutionPureState} and the excited state density matrix element from Eq. \eqref{excitedStateInput} into Eq. \eqref{PoyntingsTheoremInput} giving
\begin{equation}
\label{PoyntingInputDerivation}
\frac{\partial}{\partial Z}|\Omega_a|^2 + 2\mu \frac{\partial}{\partial T}\langle\rho_{33}\rangle = \frac{4}{\tau^2}\frac{\partial}{\partial Z }\text{sech}^2\bigg(\frac{T}{\tau} - \kappa Z\bigg) + 2\mu \bigg\langle \frac{1}{1+(\Delta\tau)^2}\bigg\rangle \frac{\partial}{\partial T} \text{sech}^2\bigg(\frac{T}{\tau}-\kappa Z\bigg).
\end{equation}
Using the definition of $\kappa$ given in Eq. \eqref{lengthScale} along with the Doppler averaging function allows us to re-write the right hand side of Eq. \eqref{PoyntingInputDerivation} as
\begin{equation}
\label{PoytingInputDerivation2}
\frac{4}{\tau}\bigg(\frac{1}{\tau}\frac{\partial}{\partial Z} + \kappa\frac{\partial}{\partial T}\bigg) \text{sech}^2 \bigg(\frac{T}{\tau} - \kappa Z\bigg) = 0,
\end{equation}
thus clearly satisfying Eq. \eqref{PoyntingsTheoremInput}.  Unlike the output Stokes pulse the input pump pulse does depend on $Z$.  In Poynting's Theorem this dependence is compensated by the corresponding time dependence of the excited state density matrix element.  Thus when written in laboratory-frame coordinates one can see that the pump pulse group velocity is reduced relative to $c$, due to the coherent excitation and de-excitation of the medium caused by the pulse.

One can similarly verify that the general solutions, given in Eqns. \eqref{PulseSolution} for the pulses and Eq. \eqref{excited_state} for the excited state density matrix element, also satisfy Poynting's Theorem given in Eq. \eqref{PoyntingsTheorem} for all $Z$ and $T$ and without pure-state assumptions.  This shows explicitly that no excitation is lost in the pulse transfer through the Raman exchange regime II, despite the initial ``obvious" implication that it was.


\section{Numerical Raman Solutions}

\begin{figure}[!h]
\includegraphics{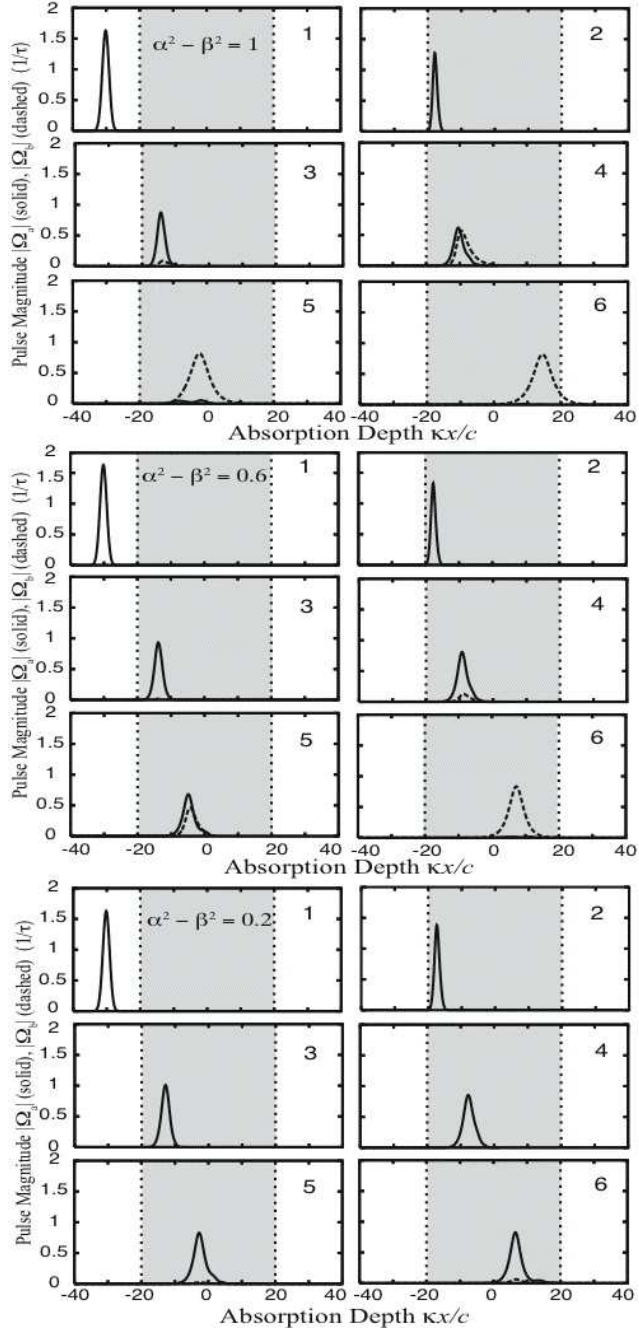}
\caption{\label{fig.Num_Raman} Snapshots of the numerical pulse solutions of  the full three-level Eqs. \eqref{DensityMatrix} and \eqref{MaxwellEquation} for $\bar\Delta = 10/\tau = 3/T_2^*$, and initial pump-probe pulses with Areas $\theta_{a}=1.3\pi$ and $\theta_{b} = 0.005\pi$. The horizontal axis is $x$ in units of $\kappa/c$, and the vertical axis is the magnitude of the Rabi frequency in units of $\tau^{-1}$.  The shaded zone indicates the location of the medium.  The solid curve is $|\Omega_{a}|$ and the dashed curve is $|\Omega_{b}|$. The top, middle, and bottom plots differ only in the way the medium is prepared. In all cases $\alpha^2  + \beta^2$ = 1, and in the three plots we have $\alpha^2 - \beta^2$ = 1.0,\ 0.6,\ and 0.2, respectively from top to bottom. The results show the emergence of an amplified probe pulse that approaches sech shape and Area $\approx 2\pi$, as in SIT.  }
\end{figure}

Our analytic solutions are clearly highly specialized. To test their broad relevance, we examine more ``normal" pulse evolution now by numerical methods and look for correspondences with prominent features of the analytic solutions.  Different medium preparations and input pulses are examined. We use gaussian instead of sech pulses and replace the infinite uniform medium by a medium with definite entry and exit faces. In Fig. \ref{fig.Num_Raman} we show three plots, each containing a sequence of six snapshots of the pulse evolution.  Each plot corresponds to a different medium preparation.  Given a medium prepared with $\alpha^2 > \beta^2$, the pump must eventually be absorbed and the probe pulse amplified, until the pump is fully depleted. It is useful to define $Z_T$ to be the ``transfer length," which is the length of the medium needed for the two pulse Areas to be equal.  We can estimate this from Eqn. \eqref{PulseAreas} by defining $\theta_b(-Z_T)=\theta_b^{\text{(in)}}$ where $\theta_b^{\text{(in)}}$ is the input area of the Stokes pulse (see Fig. \ref{fig.AreaEvol}).  Using this definition, we solve for $Z_T$ giving
\begin{equation}
\label{transfer_loc}
Z_T = \frac{1}{2\kappa(\alpha^2-\beta^2)}\ln\left(\left(\frac{2\pi}{\theta_b^{\text{(in)}}}\right)^2-1\right).
\end{equation}
For the left plot the predicted transfer location is $\kappa Z_T \approx 6$, and we see from frame 4 of the left plot that this is confirmed.  For the center plot we predict $\kappa Z_T \approx 10$, which is also confirmed. For the right plot the analytic formula gives $\kappa Z_T \approx 30$, but the last frame only shows the pulses to $\kappa Z \approx 25$ so the Stokes pulse is still weak.  We note that these confirmations are obtained despite the fact that the input pulses are gaussian:
\begin{equation}\label{numInput}
\Omega_{a}^{(in)}  = \frac{\theta_{a}}{\tau\sqrt{2\pi}}e^{-\frac{T^2}{2\tau^2}}\ \quad {\rm and}\ \quad 
\Omega_{b}^{(in)}  = \frac{\theta_{b}}{\tau\sqrt{2\pi}}e^{-\frac{T^2}{2\tau^2}},
\end{equation}
and the input pump-pulse Area is not $2\pi$.  Because we are starting with a pulse area $\theta_a < 2\pi$, the actual transfer location is slightly after the predicted location, which also explains the weak nature of the Stokes pulse in the right plot of Fig. \ref{fig.Num_Raman}.  The most efficient transfer clearly occurs when the Raman inversion, $\alpha^2 - \beta^2$, is greatest.

The I-II-III three-regime behavior we have described is apparent in all three plots in Fig. \ref{fig.Num_Raman}, and we find the same behavior for a wide variety of input pump pulse Areas with gaussian shaped envelopes and different mixed-state medium preparations. Our analytic solution for the output Stokes pulse in regime III, given by Eq. \eqref{OutputPulseSolution}, is a very good approximation to the numerical Stokes pulse.  The output pulse tails show the exponential behavior characteristic of a sech shaped pulse, and the Area of the output Stokes pulse is nearly independent of the pump pulse Area and near to $2\pi$. That is, even when two pulses copropagate, the single-pulse principles of SIT remain the strongest determiner of pulse evolution. 

\begin{figure}[h]
\begin{center}
\leavevmode
\includegraphics{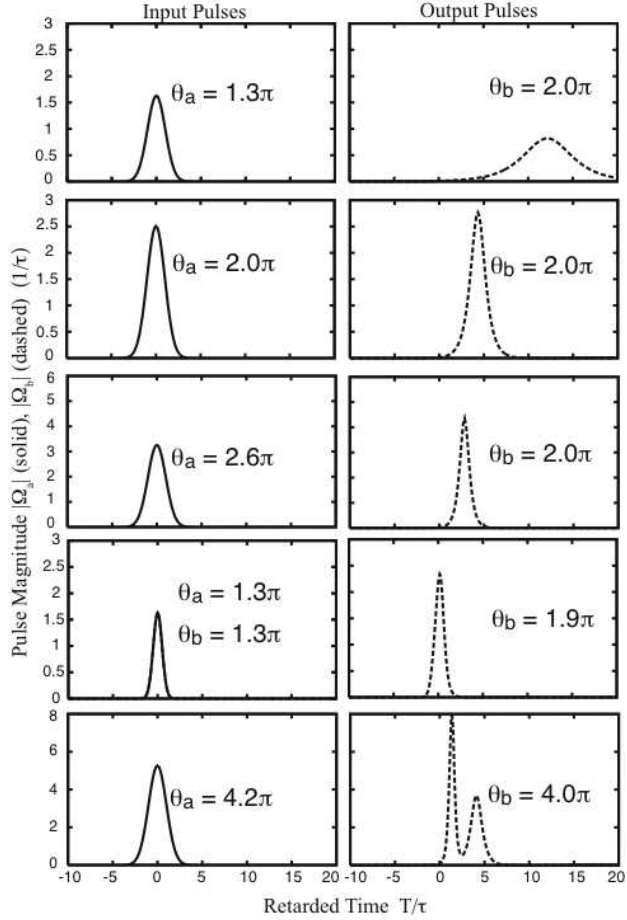}
\end{center}
\caption{\label{fig.InputOutput} Plots of numerical pulse solutions of Eqs. \eqref{DensityMatrix} and \eqref{MaxwellEquation} for ``strong" and ``weak" input Areas. Here the display is given as a function of $T$ instead of $Z$.  The pump pulses have ``strong" input Areas as indicated on the plots, and the probe pulses have ``weak" Areas $\theta_b = 0.005\pi$ in all but the 4th plot. The top frame corresponds exactly to the input and output solutions shown in the top plot of Fig. \ref{fig.Num_Raman}. The left column of snapshots shows the gaussian input pulses at the entrance face to the medium. The right column shows the output pulses at the output face after the pulses have propagated $\kappa Z = 40$ where the output Stokes pulse is well described by the sech-shaped $2\pi$ output pulse of Eq. \eqref{OutputPulseSolution} in all but the last frame.  In all frames both pump and Stokes pulses are plotted, however due to scale (or overlap as in the 4th plot) just the pump or Stokes can be visualized. The last frame shows an example of an SIT-type breakup.  All of these solutions are for media with $\alpha^2$=1 and $\beta^2$=0.  }
\end{figure}

This conclusion is ``universal" in the same sense that $2\pi$ $sech$ pulses are the ``universal" consequences of pulse propagation in two-level media if the input Area satisfies $\pi < \theta_a < 3\pi$ (of course, under the fully coherent conditions that the pulse durations are short enough to ignore homogeneous relaxation and the medium is inhomogeneously broadened). For input Area greater than this range a more complicated pulse-breakup behavior enters the picture.  To illustrate all of this concretely, we show in Fig. \ref{fig.InputOutput} just the input and output pulse shapes starting from input pulses that are gaussian with Areas $\theta_a$ indicated on the plots and $\theta_b=0.005\pi$.  In contrast to all previous plots, here we plot the pulse envelopes as a function of the retarded time variable $T$.  In the bottom frame  one can see pulse breakup because the input Area is greater than $3\pi$. In previous work with two-pulse evolution in three-level media \cite{tanno} similar breakup was reported in numerical solutions of the reduced two-level equations. Here we see that the underlying cause for the breakup is related to higher order SIT-soliton behavior in the fundamental three-level equations.  We will not pursue the study of larger-Area pulses here, but higher order analytic solutions to the fundamental equations are available for the case of resonant interaction \cite{park-shin}.

\section{Comparison to Adiabatic SRS Theory}

Our short-pulse theory of SRS retains the full three-level character of the medium, but SRS is typically studied in the far off-resonant regime where the average detuning $\bar{\Delta}$ is so large that the excited state (level 3) can be adiabatically eliminated from the density matrix equations.  The remaining two-level adiabatic MB equations have been extensively analyzed in the context of traditional Raman scattering with {\em cw} pump and probe fields \cite{raymer-etal, raymer-mostowski}.  Both ``steady state" (homogeneous decay dominant) and ``transient" (homogeneous decay neglected) analytic and numerical solutions have been found, assuming {\em cw} fields.     

Given the analytic solutions presented above, we can examine the adiabatic two-level equations in a new way, simply by comparing exact with adiabatic in the limit of large detuning. To quickly review the standard adiabatic elimination technique \cite{allen-stroud, crisp, milonni-eberly78}, we formally integrate $\dot{\rho}_{13}$:
\begin{equation} \label{rho13Eqn}
\dot{\rho}_{13} = i\Delta {\rho}_{13} + \frac{i}{2}[\Omega_a(\rho_{33} -\rho_{11})-\Omega_b\rho_{12}],
\end{equation}
with the initial condition $\rho_{13} = 0$ at $t = -\infty$, to obtain
\begin{equation}
\label{Ad_int1}
\rho_{13}(t)=i\int_{-\infty}^t e^{i\Delta(t-t^{\prime})}A(t^{\prime})dt^{\prime}.
\end{equation}
Here $A(t) = \frac{1}{2}[\Omega_a(t)(\rho_{33}(t) -\rho_{11}(t)) -\Omega_b(t)\rho_{12}(t)]$, and we have suppressed the $Z$ dependences.  We integrate Eq. \eqref{Ad_int1} by parts repeatedly to obtain the usual series 
\begin{equation}
\label{Ad_int2}
\rho_{13}(t) = -\frac{A(t)}{\Delta} + \frac{i}{\Delta^2}\frac{\partial A(t)}{\partial t} + \frac{1}{\Delta^3}\frac{\partial^2 A(t)}{\partial t^2} + \dots 
\end{equation}
In the case at hand, where each field consists of a single pulse, the ``strength" of the interactions can be estimated from the pulse Areas $\theta_a$ and $\theta_b$. For the sake of simplest estimates we can take both pulses to be roughly similar to the extent that we can denote both of their durations as $\tau$, as we already indicated in Eqn. (\ref{numInput}), for example. Then we have $\Omega_a \sim \theta_a/\tau$ and $\Omega_b \sim \theta_b/\tau$ and by extension $\partial\Omega_a /\partial t \sim \theta_a/\tau^2$, etc.   As a consequence 
\begin{eqnarray} \label{estimatedA}
A(t) & \sim & \frac{M}{\tau},\quad {\rm and}\quad \frac{\partial A}{\partial t} \sim \frac{M}{\tau^2},\quad {\rm etc.}, 
\end{eqnarray}
where
$$M  \equiv \frac{1}{2}[\theta_a(\rho_{33} - \rho_{11}) -\theta_b\rho_{12}],$$
and $M$ is clearly bounded by the magnitudes of $\theta_a$ and $\theta_b$, which will be taken of order $2\pi$ or much smaller. 

The consequence of these estimates is to turn Eqn. (\ref{Ad_int2}) into a power series in the parameter $1/\tau\Delta$. In traditional SRS this is a very small parameter because the detuning is typically large and the pulse duration is also large or even extremely large in the case of quasi-{\it cw} fields. Then only the first term is needed in (\ref{Ad_int2}), and one obtains the familiar ``adiabatic following" solution $\rho_{13} \approx -A(t)/\Delta$, equivalent to setting $\dot{\rho}_{13} = 0$ in the density matrix equations.  This lowest order solution for $\rho_{13}$ and a similar one for $\rho_{23}$ can be paired with the relation $\rho_{33} \approx 0$, which also follows from large $\tau\Delta$ values when Areas are not large. Then the three-level density matrix  equations \eqref{rhoEquations} simplify to equations for the density matrix of the two ground states alone:
\begin{subequations}
\label{reducedDensityMatrix}
\begin{align}
\dot{\rho}_{11} & = i\frac{\Omega^{(2)}}{2}\rho_{21} - i\frac{\Omega^{(2)*}}{2}\rho_{12} \\
\dot{\rho}_{22} & = i\frac{\Omega^{(2)*}}{2}\rho_{12} - i\frac{\Omega^{(2)}}{2}\rho_{21} \\
\dot{\rho}_{12} & = i\frac{\Omega^{(2)}}{2}(\rho_{22}-\rho_{11}) + i\Delta^{(2)}\rho_{12},
\end{align}
\end{subequations}
where $\Omega^{(2)}=\Omega_a\Omega_b^*/2\Delta$ is the two-photon Rabi frequency and $\Delta^{(2)}=(|\Omega_a|^2-|\Omega_b|^2)/4\Delta$ is the two-photon AC Stark shift. The same procedure simplifies Maxwell's equation \eqref{MaxwellEquation}, giving
\begin{subequations}\label{reducedMaxwell}
\begin{align}
\frac{\partial\Omega_a}{\partial Z} & = -i\frac{\mu_a}{2\Delta}(\Omega_a\rho_{11} + \Omega_b\rho_{12}) \\
\frac{\partial\Omega_b}{\partial Z} & = -i\frac{\mu_b}{2\Delta}(\Omega_b\rho_{22} + \Omega_a\rho_{21}).
\end{align}
\end{subequations}

In comparing with the analytic solutions we first expand them in a series of powers of the small parameter $1/\tau\Delta$, and then insert the lowest order contributions to those solutions into the two-level equations. One finds agreement, confirming the lowest-order adiabatic approach, but with a significant qualification. An even-handed comparison requires that the two-level equations themselves be first reduced to the internally consistent lowest adiabatic order. In the most common context, the SRS process involves amplification of a weak Stokes pulse by a strong pump pulse, in which case only the leading term in the first of Eqs. \eqref{reducedMaxwell} remains, with $\rho_{11} = \alpha^2 = constant$. This reduces the equation to one that predicts only dispersive (phase) change with no absorptive effect at all - no change in pump amplitude. It is straightforward to see that this is consistent with the exact solutions to the same order because the propagation coefficient in Eq. \eqref{lengthScale} reduces to $\kappa \approx 0$ in the $\tau\Delta \gg 1$ limit, preventing any susubstantial changes in amplitude during propagation, allowing only phase change.   

\begin{figure}[h]
\includegraphics{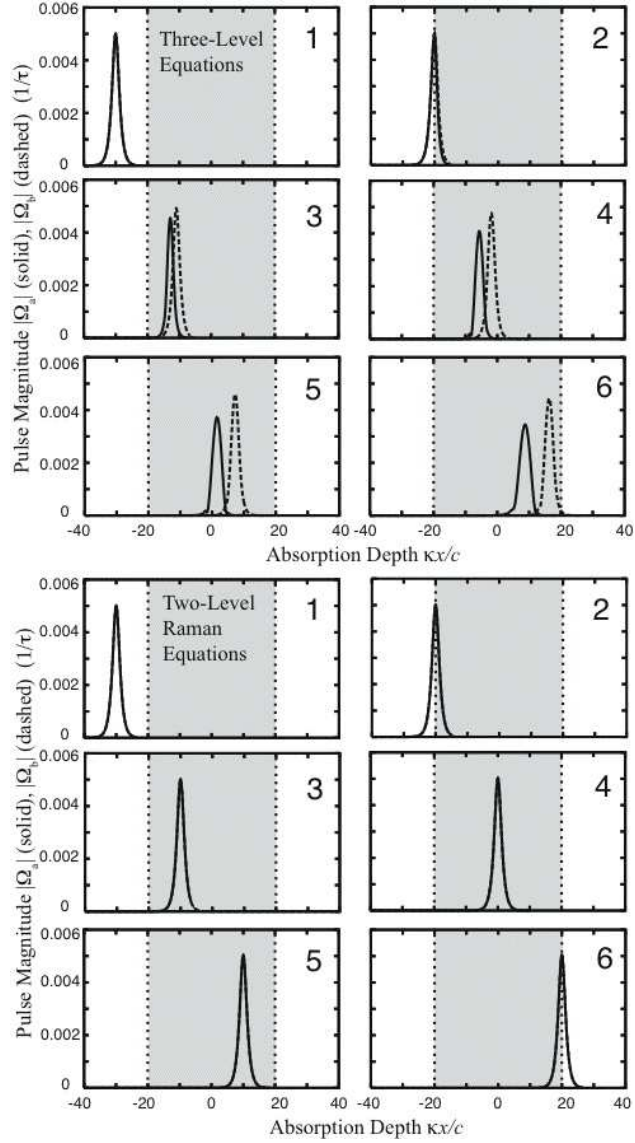}
\caption{\label{fig.Weak} \footnotesize{Snapshots of weak pulse propagation, with horizontal and vertical axis the same as similar plots.  Top six frames show plots of numerical solution to full three level equations. Bottom six frames show numerical solution to reduced two-level Raman equations. The medium in both cases is a mixed state with $\alpha^2=0.8$ and $\beta^2=0.2$.  Both input pulses are weak and sech-shaped with input area $0.005 \pi$. The three-level equations show both pulses being absorbed by the medium with differing group velocities and minimal pump-Stokes transfer as expected for weak pulses.  The reduced two-level equations cannot accurately describe the reduced group velocity and absorption effects that clearly occur in the full three-level case.  In this weak pulse example, the two-level equations cause minimal amplification of the Stokes pulse, causing the pump and Stokes pulses to appear on the plot as only a single pulse.} }
\end{figure}

\begin{figure}[t]
\includegraphics{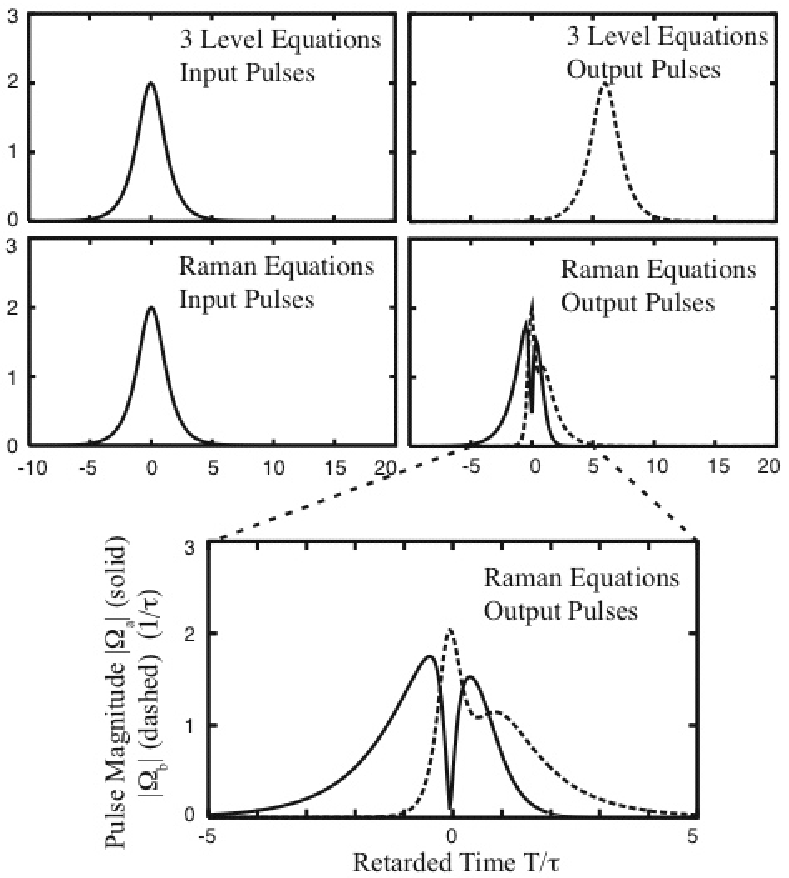}
\caption{\label{fig.3L_Raman_Comp} Evolution predicted by the complete three-level equations (top row) and by adiabatic two-level equations (second row). The left and right frames show the pulses at input and output before and after the medium (at $\kappa Z = 40$) prepared with $\alpha^2 = 1$ and $\beta^2= 0$. The horizontal axis is the retarded time $T$ in units of the pulse width $\tau$, and the vertical axis measures the product $\Omega\tau$ for each pulse.  The solid curve is $|\Omega_{a}|$ and the dashed curve is $|\Omega_{b}|$, which is hardly visible in the input plots, but is plotted. }
\end{figure}

The remaining question, therefore, is whether the SRS two-level equations can be confirmed beyond this lowest order, the order in which they were derived. What is found by comparison beyond lowest order is that the traditional two-level SRS theory does not consistently predict the short-pulse Raman transfer process developed in the preceding Section. Important details of those solutions are missing. This conclusion can be examined in numerical solutions, as in Figs. \ref{fig.Weak} and \ref{fig.3L_Raman_Comp}. We first consider injecting two weak pulses, to isolate the effects of a short pulse duration, as shown in Fig. \ref{fig.Weak}. The differences in the pulse evolution are clear, showing that short pulse propagation is not reliably reproduced.  Specifically the SRS two-level equations do not reproduce the reduced group velocity and absorption of the pump pulse.  Only for short medium lengths is the group velocity dispersion negligible, making the SRS two-level equations a valid approximation.

Further comparison of the SRS two-level equations with the three-level equations can be made by examining details of the outputs obtained when we inject two sech pulses with different Areas, one of them large and the other small: $\theta_a=2\pi$ and  $\theta_b=0.005\pi$, and  neglect inhomogeneous broadening.  In the left frames of Fig. \ref{fig.3L_Raman_Comp} we show the pulse envelopes before entering the medium and in the right frames we plot the output pulses after exiting the medium at $\kappa Z = 40$.  The first row is the solution to the full three-level equations, and the second row is the solution using the reduced equations, and the bottom figure is a zoom plot of the output pulses in the second row. In the second row one can clearly see the inadequacy of the conventional adiabatic elimination process in this case. The Stokes pulse only grows where the pump pulse has been depleted to satisfy the Manley-Rowe relation $\frac{\partial}{\partial Z}(|\Omega_a|^2 + |\Omega_b|^2) = 0$ [easily established via Eqns. \eqref{reducedMaxwell}], and the pump pulse is never fully depleted.  In addition because $\kappa \approx 0$ in the adiabatic limit, the pulse transfer occurs at velocity $c$.  In contrast, the three-level solution accurately fits the analytic transfer solution Eq. \eqref{PulseSolution}.

Incidentally, we note that the complicated pulse shape of the two-level solution in Fig. \ref{fig.3L_Raman_Comp} is similar to the early numerical results presented by Tanno, et al. \cite{tanno}, where the reduced two-level equations were solved for gaussian shaped input pulses. As a result of our comparisons we can now suggest why those earlier numerical solutions resisted clear interpretation. It appears that the pulse durations and strengths were used in combinations to make pulse Areas greater than permitted for validity of the two-level equations. In particular, we believe that the upper level should have played a role and needed to be included.

\section{Conclusions}

We have presented an analysis of the coupled Maxwell-Bloch equations governing two-pulse propagation in lambda-type media, under conditions that can be called ``short-pulse EIT". By this term we mean that the equations were solved under conditions familiar in EIT scenarios except that the pulses are taken very short compared to medium relaxation times instead of very long, and both pulses are  propagated without typical EIT pump-probe intensity relationships. This is a regime where experimental tests have yet to be undertaken, but no fundamental barriers to tests appear to be present.  

Our solution formulas for two-pulse propagation were obtained by the B\"acklund transformation method as refined by Park and Shin \cite{park-shin}, and were built on initial conditions that included incoherently prepared ground states. The solutions are complicated but they have easily understood asymptotic forms for both $x \to -\infty$ and $x \to +\infty$, although these limit forms are different from each other. This difference is compatible with stimulated Raman scattering, where initial and final stages are very different. The solutions allowed us to undertake comparisons not previously made, between propagation in pure-state and mixed-state media, and we could assess the role of the atomic ``dark state" \cite{Arimondo} in a new context. Our results show that there is essentially no role here for the dark state as $x \to \infty$, in direct contrast with the results for the same pulse preparation in pure-state media \cite{Kozlov-Eberly}.

Our analytic solutions were shown to be stable in the important sense that the output results they predict are also obtained from numerical integration of the same equations, but without needing to insist on either $sech$ shape or soliton-perfect conditions on the input. This again shows that the $2\pi$ condition derived in the McCall-Hahn Area Theorem \cite{mccall-hahn} is sufficient to determine asymptotic pulse formation in an absorbing medium. We see that the same medium, prepared in the same way throughout $[-\infty \le x \le +\infty]$, serves as a ``pump" attenuator in the early stages and as a ``probe" attenuator in the final stages. 

We took advantage of the numerical approach to make a test of the familiar reduced two-level ``adiabatic" Raman equations, and found they are not adequate for the short-pulse domain studied here.  The next term beyond the first-order approximation in the standard adiabatic series must be included for agreement with the analytic solution formulas. Of course, in going outside the $2\pi$ Area domain one encounters soliton breakup, and this is proposed as the explanation for non-smooth pulse shapes obtained in previous numerical studies.

\section{Acknowledgements}
We thank Q-Han Park for helpful discussions and correspondence.  B.D. Clader acknowledges receipt of a Frank Horton Fellowship from the Laboratory for Laser Energetics, University of Rochester. Research supported by NSF Grant PHY 0456952 and PHY 0601804.  The e-mail contact address is: dclader@pas.rochester.edu.

\end{document}